
\magnification=1200
\baselineskip=16pt
{}.
\vskip 2cm
\centerline{\bf On the Critical Temperature of Non-Periodic Ising Models}
\centerline{\bf on Hexagonal Lattices}
\vskip 2cm
\centerline{Ferenc Igl\'oi and P\'eter Lajk\'o}
\vskip 1cm
\centerline{Research Institute for Solid State Physics}
\centerline{H-1525 Budapest, P.O.Box 49, Hungary}
\smallskip
\centerline{and}
\smallskip
\centerline{Institute for Theoretical Physics, Szeged University}
\centerline{H-6720 Szeged, Aradi V. tere 1, Hungary}
\vskip 2cm
{\bf Abstract:}
The critical temperature of layered Ising models on triangular and
honeycomb lattices are calculated in simple, explicit form for
arbitrary distribution of the couplings.
\vskip 1cm
PACS-numbers: 05.50.+q, 64.60.Cn, 64.60.Fr
\vfill
\eject
In recent years we have witnessed a growing theoretical interest to understand
the
critical properties of non-periodic (quasi-periodic, aperiodic or
random) systems[1-4]. Periodic systems, which are built of regularly spaced
blocks or unit cells, are assumed to share the critical properties
of the homogeneous model. Indeed close to the critical point the
correlation length is much larger than the linear size of the
blocks, thus averaging over correlated domains the periodic system becomes
homogeneous.
This argument naturally fails, if the size of the unit cell is
diverging, i.e. for non-periodic models.
In these
systems - in the spirit of the Harris criterion[5] - one should
investigate the fluctuations of the couplings in a domain of
correlated spins. The perturbation, caused by the non-periodicity is
then irrelevant (relevant), if the local energy fluctuations are
smaller (greater) than the corresponding thermal energy.

In study the critical behaviour of specific systems it is very important to
know the position of the critical point exactly, since
with this information the
accuracy of the numerical methods is largely increased. In
two-dimensions the position of the critical point of some homogeneous
models can be deduced from duality relations[6]. The same method works
for a random bond Ising model on the square lattice, where the two
types of bonds are distributed randomly with equal probability[7],[8].
For periodic Ising models on the square lattice with unit cells of
size $m \times n$ an exact relation for the critical temperature
can be derived with the use of the Pfaffian method[9] or the transfer matrix
technique[10].
This relation is given by a matrix equation which
involves $m$-product of non-commuting $2^{n-1} \times 2^{n-1}$
matrices. Therefore in the non-periodic limit $m \to \infty$ the
critical temperature is explicitly known only for $n=1$, i.e. for
simple layered systems[11].

The homogeneous Ising model has also been solved exactly on
triangular and honeycomb lattices for a long time[12],[13]. The partition
function of the models on these hexagonal lattices are related
through a star-triangle transformation[6]. A repeated use of this
mapping is the basis of an exact renormalisation group
transformation by Hilhorst, Schick and van Leeuwen [14], which method
has also been used to study the surface critical behaviour of layered
models[15]. With this method one can also obtain numerical estimates on
the critical point of layered Ising models on hexagonal lattices,
however up to now the exact position of the phase transition temperature is
not known analytically.

In the present paper we consider this problem
and derive an explicit expression for the
critical temperature for arbitrary distribution of the couplings
both on the triangular and honeycomb lattices.
In the calculation one may directly generalize Houtappel's solution
for the homogeneous model[12]. We will, however, proceed on a
simpler way and utilize the correspondences between the square
and hexagonal lattices. Let us consider a square lattice, where
every second horizontal couplings have a special strength, denoted by
$K$ (Fig 1a). Taking the limit $K \to \infty$ the lattice becomes equivalent
to the triangular
lattice, whereas for $K=0$ one is left with a honeycomb lattice.
Thus the exact
solution of layered triangular and honeycomb lattice Ising models
can be obtained from the corresponding results of periodic
models with a $m \times 2$ unit cell on the square lattice.

The problem of periodic Ising models on the square lattice has been solved
analytically first by Hamm[9] with
the Pfaffian method and later by Hoever[10] with the transfer matrix
technique. Here we use results and notations of the second
method.

Let us consider $M \times N$ Ising spins on a square lattice. As indicated
on Fig.1a the
vertical couplings $K_j^k$ ($j=1,2,\dots M$, $k=1,2,\dots N$) are
given by:
$$K_j^k=\cases{K~~~k+j=odd\cr
               K_j~~~k+j=even\cr} \eqno(1a)$$
whereas the horizontal couplings are the same within one column:
$$\tilde K_j^k=\tilde K_j \eqno(1b)$$
The model is periodic in the horizontal direction: $K_{j+m}=
K_j$, $\tilde K_{j+m}=\tilde K_j$. In the following we consider
only purely ferromagnetic models, i.e. $K_j>0,~\tilde K_j>0$. As
mentioned before $K \to \infty$ and $K=0$ correspond to the
triangular and honeycomb lattices, respectively. Introducing the
notations
$$\tilde x_j=\rm{sinh}(2\tilde K_j)=\left(\rm{sinh}(2
L_j)\right)^{-1} \eqno(2)$$
one can write the partition sum of the model as[10],[11]:
$$Z=\prod_{j=1}^m\left( 2\tilde x_j \right)^{NM \over 2m} trace~T
\eqno(3)$$
where $T$ is the transfer matrix:
$$T=\prod_{j=1}^m T(j),~~~T(j)=T_1(j) T_2(j) \eqno(4)$$
Here
$$T_1(j)=\exp \left(L_j \sum_{k=1}^N \sigma_k^x \right) \eqno(5)$$
whereas $T_2(j)$ is different for odd and even indices:
$$\eqalign{T_2(2l-1)&=\exp\left( \sum_{k=1}^{N/2} K_{2l-1} \sigma_{2k-1}^z
\sigma_{2k}^z + K \sum_{k=1}^{N/2} \sigma_{2k}^z
\sigma_{2k+1}^z \right) \cr
T_2(2l)&=\exp\left(K \sum_{k=1}^{N/2} \sigma_{2k-1}^z
\sigma_{2k}^z + \sum_{k=1}^{N/2} K_{2l} \sigma_{2k}^z
\sigma_{2k+1}^z \right) \cr} \eqno(6)$$
and $\sigma_k^x,~\sigma_k^z$ are Pauli matrices at site $k$.

After performing the Jordan-Wigner transformation the transfer matrix is
expressed in terms
of fermion operators, which in Fourier representation reads as:
$$T=\prod_{0\le q < \pi/2} T(q),~~~T(q)=\prod_{j=1}^m
T_j(q),~~~T(j)=\prod_{0\le q < \pi/2} T_j(q) \eqno(7)$$
where the allowed values of $q$ are equidistantly spaced by $2\pi
/M$. Denoting the leading eigenvalue of the transfer operator
$T(q)$ by $\exp(\lambda (q))$, the free energy per spin $f$ is given by:
$$-\beta f={1 \over 2m} \sum_{j=1}^m \ln(2\tilde x_j) +{1 \over
2\pi m} \int_0^{\pi/2} \rm{d}q \lambda(q) \eqno(8)$$
At the phase transition temperature the free energy is singular, which is
a consequence of the singularity in $\lambda(q)$ at the border of
the Brilliouin-zone, i.e. at $q_c=0$. Our aim in the following is
to determine the position of this singularity, i.e. to calculate
the critical temperature of the system.

First, following Ref[10] we introduce the parity operator
$$P=\sigma_1^x \sigma_2^x \eqno(9)$$
which commute with the transfer operator $T(q)$, thus $T^P(q)$ acts only
on a subspace with parity eigenvalue $P=\pm 1$. For sufficiently
high temperatures the largest eigenvalue $\lambda(q_c)$ is in the
sector with $P=1$, whereas for $T=0$ it is usually in the sector
with $P=-1$.\footnote{\dag}{It could happen in frustrated systems with
ferro- and antiferromagnetic couplings, that at any finite
temperatures $\lambda^+(q_c)>\lambda^-(q_c)$.
Then the free
energy is analytical and there is no phase transition in the system.}
Since the jump in the largest eigenvalue from one
sector to the other can not be done analytically there is a phase
transition point in the system at a temperature $T_c$ where
$$\lambda^+(q_c,T_c)=\lambda^-(q_c,T_c) \eqno(10)$$
At the border of the Brillouin-zone the transfer
operator $T_j^P(q_c)$ can be easily evaluated.
It is a $2 \times 2$ matrix given by[10]:
$$T_j^P(q_c)=\left(\matrix{C_jE_j&S_jE_j\cr
                           S_jE_j^{-1}&C_jE_j^{-1}\cr}\right)
\eqno(11)$$
where $C_j=\rm{cosh}[L_j(1+P)]$, $S_j=\rm{sinh}[L_j(1+P)]$ and
$E_j=\exp\left[(K-K_jP)(-P)^j\right]$. Thus the transfer operator
is different for odd and even rows.

In the following we consider the triangular lattice, which
corresponds to $K \to \infty$. In this limit the
transfer operator has a simple structure. Keeping terms in leading exponential
order the product of two consequtive transfer
operators is given by:
$$\prod_{j'=j}^{j+1}T_{j'}^+(q_c)=\left[\prod_{j'=j}^{j+1}\exp(K) \exp(-K_{j'})
\rm{sinh}(2L_{j'})\right]
\left(\matrix{0&0\cr\rm{coth}(2L_{j+1})&1\cr}\right)+O(1) \eqno(12a)$$
and
$$\prod_{j'=j}^{j+1}T_{j'}^-(q_c)=\left[\prod_{j'=j}^{j+1}\exp(K) \exp(K_{j'})
\right]
\left(\matrix{1&0\cr0&0\cr}\right)+O(1) \eqno(12b)$$
for the $P=1$ and $P=-1$ sectors, respectively.
Similarly, in this limit any product of the $T_j^P(q_c)$-s
is given in a simple form and the eigenvalues
$\lambda^{\pm}(q_c)$ can be expressed explicitly. Then from eq(10) one
obtains the criticality condition as:
$$\prod_{j=1}^m \rm{sinh}(2\tilde K_j) \exp(2 K_j)=1 \eqno(13)$$
This remarkably simple formula is similar to the critical point
condition obtained on the square lattice for a simple layered
model, i.e. for $n=1$[11]. It is easy to check that eq(13) contains,
as a special case, the criticality condition for the homogeneous
system[12].

Next we turn to study the critical point condition for the
honeycomb lattice. Let us first consider that system which is obtained from
the square lattice on Fig 1a by removing the wavy bonds. The transfer operator
$T_j^P(q_c)$ of the system, given in
eq(11) with
$K=0$, is still a $2 \times 2$ matrix, i.e. not essentially simpler than for
the original problem on the square lattice.
Therefor there is no explicit relation, similar to eq(13), for the critical
point of
the layered honeycomb lattice Ising model which corresponds to Fig 1a with
vanishing $K$
bonds. However, if the layered structure is that shown on Fig 1c,
i.e. the vanishing bonds on the corresponding square lattice are
horizontal, then one can derive a similar relation to eq(13).
Here, instead of repeating steps of the previous calculation we make
use the star-triangle mapping[6] and transform the
result in eq(13). This transformation maps the honeycomb
lattice with bonds $p_j$, $\tilde p_j$ onto a triangular one with
lattice constants $K_j$ and $\tilde K_j$, such that
$$\exp(-4K_j)={\rm{cosh}^2(p) \over \rm{cosh}(p_j+2 \tilde
p_j)\rm{cosh}(p_j-2 \tilde p_j)}~~,~~
\exp(-4 \tilde K_j)={\rm{cosh}(p_j+2 \tilde
p_j) \over \rm{cosh}(p_j-2 \tilde p_j)} \eqno(14)$$
Substituting these relations into eq(13) one obtains the
criticality condition for the layered honeycomb lattice:
$$\prod_{j=1}^m\rm{sinh}(2\tilde p_j) \rm{tanh}(p_j)=1 \eqno(15)$$
As a special case this equation contains the criticality
condition for the homogeneous Ising model on the honeycomb lattice[12].

We note that relations in eqs(13) and (15) have been numerically verified
by the recursion method of Hilhorst and van Leeuwen[15]. Results on the
surface critical behaviour of aperiodic Ising models, obtained by
that method will be presented in a separate publication[16].
\vskip 1cm
{\bf Acknowledgement:} This work has been supported by the Hungarian 	National
Research Fund under grant No. OTKA TO12830.
\vfill
\eject
{\bf References:}
\vskip 1cm
\item{ [1]} J.M. Luck, J. Stat. Phys. 72, 417 (1993)
\bigskip
\item{ [2]} F. Igl\'oi, J. Phys. A26, L703 (1993)
\bigskip
\item{ [3]} L. Turban, F. Igl\'oi and B. Berche, Phys. Rev. B49, 12695 (1994)
\bigskip
\item{ [4]} B. Berche, P.E. Berche, M. Henkel, F. Igl\'oi, P. Lajk\'o, S.
Morgan
and L. Turban, J. Phys. A (in print)
\bigskip
\item{ [5]} A.B. Harris, J. Phys. C7, 1671 (1974)
\bigskip
\item{ [6]} I. Syozi, in {\it Phase Transitions and Critical Phenomena}, Vol.1
edited by: C. Domb and M.S. Green, (London:Academic) (1972)
\bigskip
\item{ [7]} R. Fisch, J. Stat. Phys. 18, 111 (1978)
\bigskip
\item{ [8]} B. Derrida, B.W. Southern and D. Stauffer, J. Phys.(Paris) 48, 335
(1987)
\bigskip
\item{ [9]} J.R. Hamm, Phys. Rev. B15, 5391 (1977)
\bigskip
\item{[10]} P. Hoever, Z. Phys. B48, 137 (1982)
\bigskip
\item{[11]} W.F. Wolff, P. Hoever and J. Zittartz, Z. Phys. B42, 259 (1981)
\bigskip
\item{[12]} R.M.F. Houtappel, Physica 16, 425 (1950)
\bigskip
\item{[13]} H.N.V. Temperley, in {\it Phase Transitions and Critical
Phenomena}, Vol.1
edited by: C. Domb and M.S. Green, (London:Academic) (1972)
\bigskip
\item{[14]} H.J. Hilhorst, M. Schick and J.M.J. van Leeuwen, Phys. Rev. 19,
2749
(1979)
\bigskip
\item{[15]} H.J. Hilhorst and J.M.J. van Leeuwen, Phys. Rev. Lett. 47, 1188
(1981)
\bigskip
\item{[16]} F. Igl\'oi and P. Lajk\'o (to be published)
\bigskip
\vfill
\eject
{\bf Figure Captions:}
\vskip 1cm
\item{Fig 1} Connection between the square (a) triangular (b) and honeycomb (c)
lattices. The square lattice with infinitely strong $K$ bonds (denoted by wavy
lines) is equivalent to the triangular lattice, which is related to the
honeycomb
lattice through a star-triangle transformation[6].
\bigskip
\vfill
\eject
\bye